\documentclass[
    ,final         
  ]
  {aipproc}
\usepackage{graphicx, rotating,amssymb}
\layoutstyle{8x11single}

\def\sigv{\langle\sigma v\rangle}
\def\lesssim{\buildrel < \over {_{\sim}}}
\def\gtrsim{\buildrel > \over {_{\sim}}}

\begin{document}

\title{Self-annihilating dark matter and the CMB:\\
reionizing the Universe and constraining cross sections}


\author{Fabio Iocco}{
  address={Institut de Physique Th\'eorique, CNRS, URA 2306 \& CEA/Saclay, F-91191 Gif-sur-Yvette, France}
  ,altaddress={Institut d`Astrophysique de Paris, UMR 7095-CNRS Paris, Universit\'e Pierre et Marie Curie, 98bis bd Arago, F-75014, Paris, France}
}

\begin{abstract}
I summarize the recent advances in determining
the effects of self-annihilating WIMP dark matter
on the modification of the recombination history,
at times earlier than the formation of astrophysical
objects. Depending on mass and self-annihilation
cross section, WIMP DM can reproduce sizable 
amounts of the total free electron abundance 
at z$\gtrsim$6; as known, this affects the CMB temperature
and polarization correlation spectra, and can be used to
place stringent bounds in the particle mass vs cross-section
plane. WMAP5 data already strongly disfavor 
the region capable to explain the recent cosmic
positron and electrons anomalies
in terms of DM annihilation,
whereas in principle the Planck mission has the potential to
see a signal produced by a candidate laying in that region,
or from WIMPs with thermal annihilation cross-sections 
$\sigv\sim$3$\times$10$^{-26}$cm$^3$/s and
masses with values m$_\chi\lesssim$ 50 GeV/c$^2$.

 \end{abstract}

\maketitle

\subsection{Introduction}
Observational evidence of diverse nature strongly hints
toward the existence of a matter component of the Universe,
so far undetected in the electromagnetic spectrum, but only
via its contribution to gravitational signatures.
Whereas there is quite general agreement about 
the existence of such a {\it dark} matter component, 
yet its nature is unknown.
Primordial Nucleosynthesis (BBN) and Cosmic 
Microwave Background (CMB) based arguments constrain it to
be of non-baryonic origin, and a vast amount of extensions
to the Standard Model of particles has flourished in the
literature, within which many new particles comply 
(with more or less fine tuning of theories)
with the requirements that a dark matter (DM) candidate should fulfill.

A vastly popular class of models is that of the so called
Weakly Interacting Massive Particles (WIMPs),
a typical example of which are the Ligthest 
Supersymmetric Partner or Kaluza-Klein mode, that bear 
the remarkable properties to be stable
(under conservation of R and K-parity, respectively)
and self-annihilating. Intriguingly, the
self-annihilation rate arising ``naturally'' for
such candidates being subject only to weak interactions,
would produce a relic density -if they are to produced thermally
in the early Universe- compatible with that of
DM in a $\Lambda$CDM Universe.
This has often being vividly referred to as ``WIMP miracle'',
and the reader is addressed to recent reviews 
\cite{DMrevs}, for a more detailed 
discussion of the topic.

Many efforts have been dedicated in the last years to
address the modeling of direct and indirect signatures
that such class of particles should leave. Particularly interesting
are indirect signatures of astrophysical nature, as if on one hand 
the existence of a feature hardly explainable within 
a standard astrophysical scenario would be another
evidence in favor of the existence of DM,
on the other its absence allows to put constraints
in the DM model space, ruling out those candidates 
that should have left the signature.
The recent observation of a peculiar rise in 
the positron fraction (at energies
1.5GeV$\lesssim$E$_{e^+}\lesssim$100GeV) by
the PAMELA collaboration \cite{Adriani:2008zr}, 
as well as the one of an electron {\it and} positron spectrum 
(at energies 20GeV$\lesssim$E$_e\lesssim$1TeV)
inconsistent with standard galactic propagation models 
by HESS and FERMI \cite{Abdo:2009zk,Collaboration:2008aaa},
have received a wide range of interpretations both
in terms of astrophysics outside the realm of
the simple ``vanilla'' models, and DM annihilations or decays.
The focus of these proceedings is the class of 
DM annihilation interpretations, and
I address the readers to the vast literature appeared 
since the PAMELA data were released; for a review
of astrophysical classes of models see e.g. \cite{Serpico:1900zz}
and references therein.

The properties required from a self-annihilating
particle in order to reproduce the observed feature make the 
possible candidate a rather ``exotic'' one, with respect to
the standard WIMP scenario. Above all, is that the normalization
of the signal should be orders of magnitude bigger than
the one produced by a self-annihilation rate able to reproduce
the correct relic abundance of WIMPs, as mentioned above.

Within the frenetic search for a candidate able to
explain the excess in terms of DM, huge advances have been
done in the field of model constraining, and the re-discovery
of properties and phenomena for long time ignored.
 
In these proceedings I summarize the 
main findings and advances done in using
the CMB spectra in order to constrain self-annihilation
cross-sections and masses of WIMPs, obtaining some among
the strongest constraints of astrophysical nature.
As I will argue later, in fact, typical astrophysical constraints
come from local objects, and assumptions about the
DM density field 
need to be done (even within an assigned cosmological scenario),
in addition to the complicated astrophysics involved. 
An exquisite example is the propagation of
energetic antiprotons generated by DM annihilation,
the prediction of whose abundance at Earth is affected  both
by the uncertainties on charged particle propagation in
in the Galaxy and the galactic DM halo profile, that produces the 
source signal.
Whereas a feature observed in the CMB spectra would definitely
be a more indirect signature of annihilating DM than that of
a galactic one, 
I will summarize how the signal depends only on very 
well known astrophysics and only on the
assigned cosmological scenario, thus being unplagued by 
local Universe astrophysics uncertainties and
constituting an exquisite tool for constraints.

\subsection{Self-annihilating DM and energy injection into the IGM}
Before the formation of gravitationally bound structures, the 
DM density field can be approximated by a smooth, diffuse 
one\footnote{The presence of inhomogeneities does not mine the validity 
of the argument, and only make the results more conservative.},
and the annihilation rate per unit volume, {\it A}(z) reads:

\begin{equation}
\label{AnrateselfDM}
A(z)= \frac12 \rho^2_c \Omega^2_{DM} (1+z)^6  \frac{\sigv(z)}{m^2_\chi}
\end{equation}

with $n_{DM}(z)$ being the relic DM abundance at a given redshift $z$,
$m_\chi$ the mass of the dark matter particle, $\Omega_{DM}$ the 
cold dark matter fraction, $\rho_c$ the critical density of the Universe today, 
and $\sigv$(z) is the effective self-annihilation rate
which for the sake of generality here we assume to depend on the redshift $z$
(see the Section on constraints to the ``Sommerfeld'' enhancement).
The total energy 2$m_\chi$c$^2$ produced in the annihilation will however
be only partially injected into the thermal gas -to which I will refer 
in the following as Inter Galactic Medium (IGM-although improperly
as galaxies have not yet formed at the redshifts relevant for this
process): 
part of the high energy shower
produced in the annihilation will in fact not interact with the thermal gas
and stream freely through the Universe. 
Under the so called ``on-the-spot''
approximation, consisting in the assumption that the particles
failing to interact with the IGM on--the--spot (namely within a short fraction
of the Hubble time at the moment they are produced), do not interact with
the thermal gas anymore, the energy deposited at any given
time will actually only a fraction $f$(z) of the one produced, 
bearing an energy injection rate per unit volume:
\begin{equation}
\label{EnrateDM}
\frac{d E}{d t}(z)= f(z) A(z)=f(z)\rho^2_c c^2\Omega^2_{DM} (1+z)^6  \frac{\sigv(z)}{m_\chi},
\end{equation}

where $f(z)$ depends on the spectrum and characteristics of the primaries produced by
the DM annihilation, and ultimately on the nature of the DM particles itself.

The energy injection in the thermal gas, which ultimately determines
the evolution of the IGM temperature and ionization fraction, is 
therefore regulated (in this formalism) by only one DM--related parameter:
\begin{equation}
 p_{ann}(z) \equiv f(z) \frac{\sigv (z)}{m_\chi} .
\label{pann}
\end{equation}

It is crucial to remark that the fraction $f$(z) depends only on
very well known high energy astrophysics processes: mainly Inverse Compton
scattering of energy electrons and positrons over CMB,
photoionization of hydrogen and helium (effectively the only constituents of
the high redshift gas) and pair production on CMB by high energy photons.
The ultimate value of $f$(z) does therefore depend on the
composition of the annihilation shower, and eventually on the
DM candidate itself.
At high redshift (150$\lesssim$z$\lesssim$1100) the IGM is completely
opaque at energies below the keV (see for instance Fig. 2 in 
\cite{Slatyer:2009yq}), therefore once the primary particle energy 
has been degraded down to this scale, the remaining cannot escape
the IGM anymore, thus contributing to its heating and ionization, see later.
The problem of obtaining $f$(z) is thus reduced to compute the fraction
of primaries that can cool from the GeV/TeV (the scale of a typical
WIMP candidate mass) down to the keV within few Hubble times at the
relevant redshift.
In \cite{Slatyer:2009yq}, it has been recently dealt with the problem of
energy deposition in the high redshift thermal gas from very energetic 
particles: the authors studied the interaction of different classes of high
energy primaries with different spectra and their absorption by the IGM 
throughout the evolution of the Universe. 
Using this information it is possible to reconstruct the effective fraction $f(z)$ 
for any given WIMP DM model by knowing its original primary branching ratios in different
baryonic species; yet it is worth stressing that the remarkable advantage of the
formalism proposed is its complete model-independence in the CMB analysis, provided the
on--the--spot approximation is valid, see \cite{Slatyer:2009yq}.

Once part of the initial energy due to the annihilation has been degraded
down to the keV scale, the effects of such a low--energy, yet non thermal
component are equally well known: in \cite{ShullVanSteen1985},
the authors  showed that 
the final effects are to provide {\it (i)} ionization, {\it (ii)} heating 
and {\it (iii) }Ly--$\alpha$ excitation
of the thermal gas, the details and final ripartition of the three processes
eventually depending only on the temperature and original ionization fraction
of the affected gas, and recently \cite{Valdes:2008cr} provided more
accurate estimates in light of detailed MonteCarlo simulations.

Eventually, the behaviour of the high redshift thermal gas in 
presence of DM annihilation is  well posed:
an additional heating/ionization/Ly--$\alpha$ excitation source,
regulated by Eq. \ref{EnrateDM} is added, and the cosmological 
evolution of the gas under its effects can be followed.
Many authors have modified the publicly available code 
RECFAST, which computes the properties of the evolving
thermal gas in a cosmological context, by taking into account
additional sources of ionization; the reader is addressed
to \cite{ModifRec} for a non-conprehensive list of
references of the latest authors dealing with the problem.
For sake of completeness, it is worth mentioning here
the recent work of \cite{Valdes:2009cq},
in which the authors self-consistently compute the 
amount of energy deposited in the three final channels
by DM annihilation starting from the high energy cascade,
and for several initial primaries, without using
the two-step (with the break-up at keV scale) 
approach previously described here.
In \cite{Grin:2009ik}, 
the authors have presented a new numerical code
for the computation of the recombintation
of thermal gas, 
including the effects of high-$n$ states of
the hydrogen atom on the properties of the gas
(neglected for instance in RECFAST),
thus highly enlarging the precision of
the high $l$'s CMB spectrum. 

\subsubsection{The effect of structure formation}
\label{structform}
The formation of gravitationally bound structures provides a ``boost'' to the
annihilation signal: since scatterings depend on the square of
the density field, the clumping of DM particles into haloes enhances the
annihilation rate, as $\langle \rho(z)^2 \rangle \geq \langle \rho(z) \rangle^2$,
the average being performed over the entire Universe\footnote{Notice that
this argument does not apply to isotropic signals that depend linearly on the density field 
--as for instance that of decaying DM.}.
In \cite{NatarajanSchwarz}, the authors 
first computed the effects of clumped annihilating DM 
onto high redshift thermal gas evolution and CMB observables, 
and a similar analysis has been carried on for  $\sigv$=$\sigv$(z)
by \cite{Belikov:2009qx}.
In \cite{Huetsi:2009ex,Cirelli:2009bb}, the authors have 
recently dealt with
the same problem including also the smooth density field,
and recognizing the leading effect of the latter with
respect to the Recombination history.

The reason of the almost negligible contribution of DM annihilating within
structures to the ionization of the IGM gas is to be serched again in the 
transparency function of the Universe, Fig. 2 in \cite{Slatyer:2009yq}:
at the time the bulk of structure formation starts taking place --z$\lesssim$50--
the Universe has become almost completely transparent to high energy
particles. This means that primaries produced at typical energies
1MeV$\lesssim$E$_0 \lesssim$1TeV at z$\lesssim$50 do not interact
with the IGM, thus not depositing energy into it (in the formalism used
in this paper -- $f$(z$\lesssim$50) $\sim$0 ).

This occurrence has two interesting implications:
the first, of extreme conceptual relevance is that the 
effects induced on CMB by annihilating DM come
{\it only} from the smooth, diffuse density field at high
redshift. This means that (within an assigned cosmology)
the signal is completely unaffected by uncertainties typically
associated with quantities such as the halo profile, the concentration
parameter and the halo minimal mass. Therefore, any constraint
obtained from a method which makes use of this signal is
entirely free by the uncertainties that plague other methods,
e.g. galactic multimessenger, or gammas from dwarf spheroidal galaxies.

The second point is that the high energy photons produced
at redshift z$\lesssim$50 can stream, mostly unaffected until
today, thus constituting the ``bright side'' of the 
phenomenology so far described.  However,
since the diffuse extragalactic signal at Earth (even from high redshift),
is due to the structure component of the DM density field, it will
be affected by all the uncertainties of which one can so conveniently
get rid off in the CMB signal approach, see e.g \cite{Bergstrom:2001jj}.

\subsection{Effects on CMB observables}
\label{CMBobs}

\begin{figure}
  \includegraphics[height=.3\textheight]{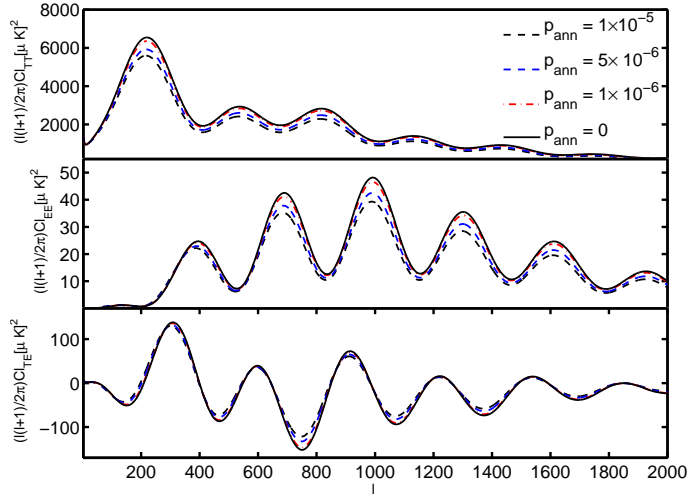}
  \caption{Modification of the CMB correlation spectra
  for different values of $p_{ann}$, from \cite{Galli:2009zc}.}
  \label{CMBspectra}
\end{figure}

The effects of an additional energy source at high redshift on the CMB observables
are of different nature, depending on the time of energy deposition;
here I limit myself to summarize briefly the phenomenology
that causes them, pointing
the reader to the main literature discussing the physics in details.

Three main time interval can be identified for the problem:

{\it i)} CMB does not provide any information about the processes 
taking place in the Universe above z$\gtrsim$10$^6$, so
DM annihilations in this time range are to produce signatures
on other observables, if any (for instance the effects on BBN,
see \cite{BBN-DM} for a recent review and a dedicated study);

{\it ii)} at 2.1$\times$10$^6\gtrsim$z$\gtrsim$1100,
during the formation of the CMB blackbody spectrum,
the energy provided by DM annihilating in the still ionized 
thermal gas is completely absorbed (i.e. $f$(z)$\sim$1).
However, since a complete thermalization of the keV 
residual photons is not possible due to the (in)efficiency of
photon non-conserving processes, distortion to the Planck spectrum
are in principle left as signature of DM annihilation;
\cite{IllioSyun:1975} and  \cite{McDonald:2000bk} have described 
in detail these processes;

{\it iii)} at 1100$\gtrsim$z$\gtrsim$150, 
the CMB blackbody spectrum is already formed;
however, the existence of a non-zero ionized fraction
will introduce thermal Syunyaev-Zeldov'ich 
effect of the residual electrons over the 
CMB photons. The TT and 
especially the TE and EE correlation spectra are 
sensitive to the distribution of free electrons between
the Recombination surface and today, and 
departure from a ``standard'' recombination history
can in principle be detected with accurate enough surveys.
In \cite{Padmanabhan:2005es}, the authors thoroughly describe
the physical processes and the distortion of the CMB 
temperature and polarization spectrum.
Here is worth recalling that
distortions in the temperature spectra due to additional thermal 
free electrons in the gas are almost entirely degenerate with the
power spectrum, as the thickening of the last scattering surface,
reduces the signal on smaller angular scales.
On the other hand, the introduction of free electrons 
after Recombination (and well
before the onset of an astrophysical Reionization) will
permit Thomson scattering of the local quadrupole of 
the temperature distribution, generating a
polarization signal on small angular scales 
(as opposed to the one created by astrophysical Reionization
at z$\lesssim$11, visible in the TE cross-correlation spectra
at angular scales of $l\lesssim$10). In Figure \ref{CMBspectra}
the TT, TE, EE correlation spectra are shown for different
values of $p_{ann}$, and for comparison with a 
``standard'' case without DM annihilations; for details
on the cosmological paramaters adopted, see the
original paper \cite{Galli:2009zc}.
\subsection{Constraining self-annihilation cross sections}
\label{constrsigv}

\begin{figure}
  \includegraphics[height=.3\textheight]{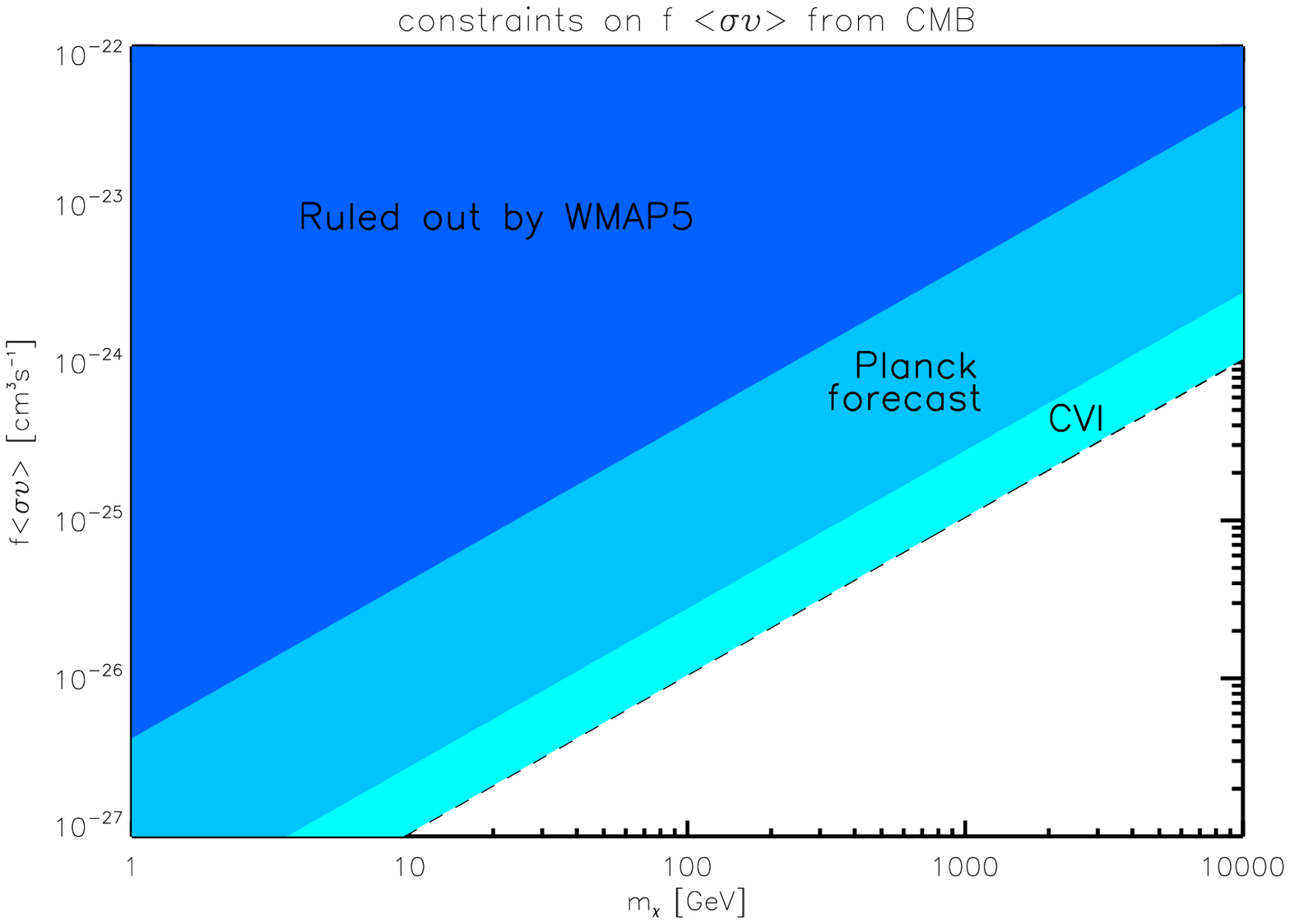}
  \caption{Constraints in the $m_\chi-f\sigv$ plane, from \cite{Galli:2009zc}.}
  \label{fsigvconstr}
\end{figure}

The existence of the signatures in the CMB spectra 
described until now
permits the possibility to look for evidence of additional
ionization induced by Dark Matter annihilating into
the thermal gas at high redshift.
In the absence of a signal, an upper limit can be put on the
parameter $p_{ann}\equiv f \sigv $/m$_\chi$,
the only free one from which signatures depend.

Although the blackbody spectrum measurement from 
COBE FIRAS spectrometer is extremely accurate \cite{Fixsen:1996nj},
not even its sensitivity allows us to draw interesting constraints
on self-annihilation cross sections for particle in the mass range of
GeV/TeV.
The analysis carried on in \cite{McDonald:2000bk}, 
does in fact show that by adopting a $f$(z)=1 throughout
the interested range of redshifts (as appropriate in this case
as the dense and ionized gas of non-yet recombined gas
is optically thick to the high energy primaries produced by DM annihilation), 
upper limits are of the order 
$\sigv\lesssim$10$^{-21}$ (/10$^{-23}$/10$^{-19}$) cm$^3$/s
for a 100GeV(/1GeV/10TeV) mass WIMP, when
assuming a $\sigv$ constant throughout the relevant redshift range.
A similar analysis has been recently performed by \cite{Zavala:2009mi},
in which the authors have studied the constraints in the parameter space of
the Sommerfeld enhancement (see the following Section),
and consequently a $\sigv$ which is a function of redshift; the results
of such analysis do not differ quantitavely if recast in terms of
efficient annihilation cross-section.

The study of the signature in the temperature and polarization
spectra gives more interesting results. 
For different values of $p_{ann}$(z) (and therefore of the
 free-electron history)  the corresponding TT, TE and EE
 cross-correlation spectra can be derived, and confronted with 
 the observed power spectra. Constraints on $p_{ann}$(z) can be 
 derived with typical statistical methods, usually
 by running a full Monte-Carlo Markov Chain analysis
 over a whole set of parameters (and including
 $p_{ann}$) and then marginalizing over the cosmological ones.
A similar analysis has been recently performed in \cite{Galli:2009zc},
under the assumption that $\sigv$ and $f$ are constant in z; the first is a 
completely justified assumption even in a Sommerfeld enhanced scenario
(see following Section), whereas a fairly good one for $f$, which is not a strongly 
varying function of the redshift $z$ in the range of interest 
(150$\lesssim$z$\lesssim$1100), see Figure 4 in Ref. \cite{Slatyer:2009yq}.

The analysis in \cite{Galli:2009zc} shows that the WMAP5 data 
can place quite strong constraints on the allowed upper limit
for $f\sigv$ if the mass of the WIMP particle is taken to be 
1GeV$\leq$m$_\chi\leq$1TeV, 
i.e. $f\sigv\leq$ 4.3$\times$10$^{-25}$ cm$^3$/s for m$_\chi$=100GeV
at 95\% confidence level, see plot in Figure \ref{fsigvconstr}.
In order to convert this into an effective upper limit
on the self-annihilation cross-section, one needs to choose
the (dominant) annihilation channel of the WIMP.
For instance, if one wants to confront with a WIMP
annihilating mainly into electrons and positrons, such
as the ``leptophilic'' models usually invoked in order to 
explain the PAMELA positron excess in terms of DM annihilations,
one finds that a good estimate for $f$, averaged in the $z$ range 
of interest, is $f\sim$0.5, Figure 4 in Ref. \cite{Slatyer:2009yq}.
This bears a constraint of $\sigv\leq$8.6$\times$10$^{-25}$ cm$^3$/s 
for a WIMP mass m$_\chi$=100GeV, at 95\% confidence
level using the WMAP5 data.

Although this value is about one order of magnitude higher
than the benchmark $\sigv\sim$3$\times$10$^{-26}$ cm$^3$/s
usually considered (in order to obtain a thermal freezeout
able to reproduce the observed cold DM relic abundance),
it is competitive with the ones obatined by galactic multimessenger
(see for instance \cite{Pato:2009fn}, and Pato et al. 09 in these proceedings) 
and unaffected by the
local astrophysics uncertainties, as previously summarized 
in the Section on structure formation.
These constraints are extremely interesting in the light
of Sommerfeld enhancement arguments, see following Section.

It is also extremely interesting to notice that a forecast 
based on mock data shows that Planck's 
sensitivity in the TE cross-correlation signal
(expecially at high $l$'s, 150$\lesssim l \lesssim$1600) will permit 
-in the case of {\it non}-detection of discrepancies from
the ``standard'' ionization history- to place constraints
of about one order of magnitude stronger than WMAP5 ones.

\subsubsection{Constraining the Sommerfeld enhancement}
\label{SommConstr}
In presence of a long range interaction, and 
at low relative velocity of the interacting particles,
the perturbative approach usually employed to
compute the self-annihilation and scattering cross section
breaks down, and high order terms can not be neglected anymore.
If DM self-interacts via the exchange of gauge bosons,
the introduction of a new effective potential must 
be taken into account; this was originally done by Sommerfeld,
\cite{Sommerfeld}, who found a 1/$v$ enhancement of the cross
section for long range interactions, here $v$ being the relative
velocity between the interacting particles, and recently
considered in the context of DM annihilations, e.g. \cite{newSommerfeld}.
The Sommerfeld enhancement has the intriguing characteristic to
preserve the self-annihilation cross section when WIMP
particles are thermally produced in the early Universe,
and $\beta\equiv v$/c $\sim$ 0.3, whereas the cross-section
should start being enhanced when $\beta \lesssim$ 10$^{-3}$/10$^{-4}$,
the latter values depending on the type of interaction and
the ratio of the DM vs the gauge boson mass in many models.
In the wake of the search for a DM interpretation of the PAMELA
signal, the Sommerfeld enhancement has received renewed attention,
as it could provide the ``boost'' needed to bring up the
signal at the correct normalization. However, 
the local galactic velocity dispersion, $\beta\sim$10$^{-3}$/10$^{-4}$
is too low to provide entirely the needed enhancement, and in many models
one needs to invoke also the contribution from substructures,
which are virialized to a smaller velocity, \cite{Lattanzi:2008qa}.
Before structure formation, and after kinetic decoupling from
the thermal gas, the thermal history of DM is described by
an adiabatic cooling in which T$\propto$z$^{-2}$;
for typical WIMP candidates with masses in the range of 
interest, at Recombination $\beta \sim$10$^{-8}$, 
and it keeps decreasing. 
Therefore, if Sommerfeld enhancement applies at all to
the model one is studying, it {\it must} be active during 
the phases relevant to affect the TT, TE and EE correlation spectra,
and the efficient $\sigv$ is a Sommerfeld enhanced one.
In many models the Sommerfeld saturates at a maximum value,
when $\beta$ drops below a given threshold, which depends on the
model, but is usually such that $\beta\geq$10$^{-8}$ for most models;
this also guarantees that the approximation of a constant
$\sigv$ is valid, and its value the Sommerfeld saturated one. It
is worth noticing that in presence of Sommerfeld enhancement the effective
self-annihilation cross section would {\it decrease} at the formation
of structures, as a consequence of the virial heating of DM. This would
even strengthen, if necessary, the argument for the negligibility
of such contribution.

It follows that the constraints derived with the method described in
the previous section can be applied ``tout court'' to models in which
Sommerfeld enhancement is present. A summarizing plot with 
the constraints in the m$_\chi$-$\sigv$ plane is Fig. 5 in \cite{Galli:2009zc},
where the effective Sommerfeld enhanced $\sigv$ (for an assigned set
of parameters, and $\beta$=10$^{-8}$) is shown.

\subsection{Discussion}
WIMP DM from the smooth, diffuse density field
annihilating at redhisfts 150$\lesssim$z$\lesssim$1100 
contributes as an additional source of ionization
of the thermal gas;
such an altered ionization history can leave a characteristic
imprint on the CMB temperature and polarization spectra.
The non-detection of a discrepancy from a standard
recombination history at 95\% confidence level in
the WMAP5 data allows us to place strong upper
limits on the normalization of the annihilation rate,
which in a convenient formalism can be parametrized by 
$p_{ann}$=$f\sigv$/m$_\chi$, which
in turn can be made model-dependent by choosing
the correct value of $f$ depending on the WIMP model and its 
annihilation channels.

Eventually, this result can be cast in the form of
exclusion plots in the WIMP mass vs self annihilation
cross section, m$_\chi$-$\sigv$ plane.
The constraints obtained, which are {\it not} plagued
by uncertainties correlated to structure formation
history or halo density profile, apply to the
self-annihilation cross section effective at the time
the distortion of the CMB spectra are generated, namely
$\sigv$(150$\lesssim$z$\lesssim$1100); this remark,
redundant within a ``vanilla'' cold dark matter scenario
thermally produced with an s-wave annihilation
cross section (since $\sigv$ is constant at any
place and time in the Universe), becomes relevant when 
dealing with more involved scenarios, such as the ones
in which a Sommerfeld enhancement is active at low relative
velocities of the DM particles; or for instance the one of
non-thermally produced DM, which then
decays (before Recombination) into the stable partner, 
which in turn self-annihilates;
a physical scenario being for instance the
mSUGRA in which a Gravitino is
the Next to Lightest Supersymmetric Partner,
decaying into the LSP stau before BBN, see e.g \cite{Steffen:2008qp}.

In principle, the Planck mission has the potential to detect 
any alteration of the recombination history, produced by annihilating
DM  with a normalization down to 
$f\sigv$=2.6$\times$10$^{-26}$cm$^3$/s for a m$_\chi$=100GeV .
It is worth stressing that these constraints are obtained within 
the assumption that annihilating DM is the only 
ionization source in addition to standard 
processes\footnote{This makes the obtained constraints
even stronger: an additional exotic ionization source
would leave less room for DM, and therefore lower the
allowed value of $p_{ann}$.},
and within a standard $\Lambda$CDM model.
The possible detection of a discrepancy from the
standard recombination history from the Planck satellite,
could not be immediately taken as evidence for DM
annihilating at high redshift, although would
definitely be {\it compatible} with it. More convincing
would be the finding of a peculiar feature imprinted
on the TE spectrum by DM annihilation, with respect
to other possible exotic ionization sources. 
Such a characteristic
feature, and the possibility to disentangle it from
other ionization sources has been recently object 
of the study by \cite{Chluba:2009uv};
it remains that for a m$_\chi\sim$100GeV thermally
produced WIMP the normalization of the signal 
would be too low to permit such a discrimination
even for Planck.
It is however worth to remark that the signature
left by WIMPs with masses m$_\chi\lesssim$50GeV,
a thermal cross-section $\sigv$=3$\times$10$^{26}$cm$^3$/s,
mainly annihilating into leptons is within the reach of Planck,
and that a non-detection would quite ultimately rule out
classes of model with these characteristics.

\subsubsection{The electron optical depth $\tau_e$}

\begin{figure}
 \includegraphics[height=.3\textheight]{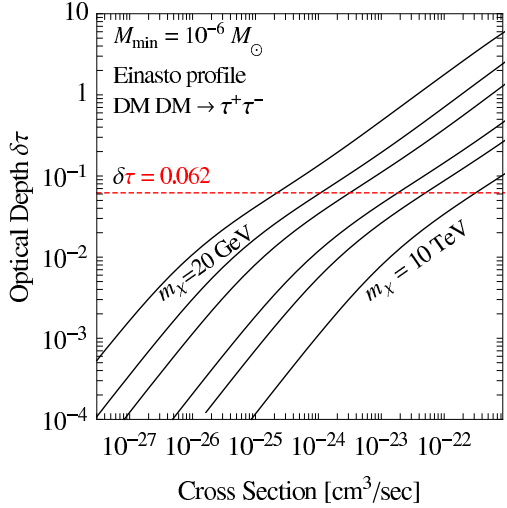}
  \caption{Contribution to the residual $\delta \tau_e$
  from different masses (lines) as a function of $\sigv$.
  Annihilation channel in $\tau^+$-$\tau^-$, see text for details;
  from \cite{Cirelli:2009bb}.}
  \label{tauplotCirelli}
\end{figure}

Such strong constraints come from the fact,
as shown by  \cite{Huetsi:2009ex,Cirelli:2009bb},
 that WIMP DM is able to produce a sizable fraction
of the electron optical depth $\tau_e$
which is formally written as:
\begin{equation}
\tau_e=- \int n_e(z)\, \sigma_{\rm T} dz ,
\label{taudef}
\end{equation}
n$_e(z)$ being the fractional abundance of free electrons
at redshift $z$ and $\sigma_{\rm T}$ the Thomson scattering
cross-section.
Let us define the residual $\delta \tau_e$ as
the integral of Eq. \ref{taudef}
between z=6 and z=700, namely the fraction
of $\tau_e$ that can be produced by self-annihilating
DM before the Universe becomes completely ionized,
and disregarding the ``recombination tag'' between 
$z$=700 and $z$=1000 (thus avoiding to
include in our computation free-electrons
from the last moments of standard Recombination,
which does in fact extend a little beyond $z$=1000).
In Figure \ref{tauplotCirelli} is shown the fraction
of the $\delta \tau_e$  that can be produced by
WIMPs of different masses (each line in the plot
representing a different mass value m$_\chi$), 
annihilating with running $\sigv$ into a $\tau^+$-$\tau^-$ channel.
The residual upper limit to 
$\delta \tau_e$ =0.062 is defined as the difference between the observed
$\tau_e$=  0.084 $\pm$ 0.016 (taking the upper limit $\tau_e$=0.10)
 by WMAP5, and Eq. \ref{taudef} integrated
 between z=0 and z=6, assuming a
completely ionized medium between today and z=6,
thus being a measure of the free electrons in the Universe 
between the recombination
and the $z$=6 surface, this latter redshift being chosen as the one
by which the entire Universe is considered to be completely
ionized, by arguments based on the absence of Gunn-Peterson
trough in the emission lines of distant quasars;
see e.g Section 1 in \cite{Cirelli:2009bb} 
for details and a thorough discussion.
It is important to stress that the choice of a particular
halo profile, concentration parameter and minimal mass 
does not affect the results at all,
as proven in \cite{Huetsi:2009ex,Cirelli:2009bb}
and discussed previously here, 
as the dominant contribution to ionization 
comes from times earlier than structure formation.

It is also interesting to notice that the use of the quantity
$\tau_e$ is well posed in a scenario where most of
the free electrons in the Universe are produced
at late redshift, namely a standard astrophysical Reionization
case, with all the free electrons produced at z$\lesssim$11;
its use starts however to be misleading in a scenario 
where the contribution to the total free electron fraction 
produced at high redshift, is non-negligible.
In particular, self-annihilating DM 
would contribute solely before the formation of
astrophysical sources, and leave smaller room
for a distinct signature in the polarization spectra 
at $l\lesssim$20, whereas the total number
of free electrons should be conserved.
Within the context we have presented, a 
study of the free-electron abundance as a function
of redshift is therefore the only well-posed way to proceed,
and such an attempt to study the differential contribution to
the canonical $\tau_e$ from different canonical sources
(including e.g. relic ionization fraction from
incomplete Recombination etc.) 
has been attempted in \cite{ShullVenka}.

\subsection{Conclusions}
WIMP dark matter self-annihilating into
standard model particles can contribute
to the ionization of the thermal gas at high redshift
(1100$\gtrsim$z$\gtrsim$150), and the 
altered free electron fraction can leave characteristic
imprints on the Cosmic Microwave Background 
TT, TE and EE correlation spectra.
These signatures have already been searched for in the 
WMAP5 data, and their absence can lead to the exclusion
of WIMPs with thermal cross sections and masses
m$_\chi\lesssim$3GeV, the actual value depending
on the nature of the primaries in which the DM annihilates into.
WMAP5 data also permit the exclusion of some of the 
most extreme models explaining the PAMELA positron
excess in terms of WIMP DM annihilation, as well as the
region that could explain PAMELA {\it and} the FERMI/HESS
electron excesses altogether.
In the absence of additional high redshift ionization signal 
in the Planck data, it will be possible to completely rule out
a vast region of the PAMELA DM interpretation space, and
probe WIMP DM with a thermal cross secton 
$\sigv\sim$3$\times$10$^{-26}$cm$^3$/s for masses
m$_\chi\lesssim$ 50GeV.
These constraints depend only on the chosen cosmology
and are unaffected by uncertainties on the structure formation
scenario and on DM halo profiles.

\subsubsection{Acknowledgments}
These proceedings are based on the experience
I have accrued during the preparation of 
\cite{Cirelli:2009bb, Galli:2009zc},
and they would have never been possible
without the collaboration and constant discussion
with all of my collaborators in this field. I am grateful to
D.~Cerde\~ no and A.~Ferrara 
for fruitful conversations and valuable comments.

\end{document}